\documentclass[aps,jmp,amsmath,amssymb,reprint]{revtex4-1}
\usepackage{amsmath, nccmath}
\usepackage{mathastext}   
\usepackage{dcolumn}
\usepackage{bm}
\usepackage{bbm}
\usepackage{graphicx}
\usepackage{subfigure}
\usepackage{sublabel}
\usepackage{physics}
\usepackage{csquotes}
\usepackage[english]{babel}
\usepackage[usenames, dvipsnames]{color}
\usepackage{hyperref}
\usepackage{hyphenat}
\usepackage[normalem]{ulem}
\usepackage{newunicodechar}
\newunicodechar{̈}{\"{}}
\usepackage{newunicodechar}
\newunicodechar{̆}{\u{}}  
\usepackage[utf8]{inputenc}
\usepackage{romannum}
\usepackage{float}
\usepackage{mathtools}
\usepackage{placeins}
\usepackage{verbatim}
\usepackage{appendix}
\usepackage{xcolor}
\usepackage[most]{adjustbox}
\usepackage{tabularx}
\usepackage{supertabular}
\usepackage{appendix}

\begin{document}
\bibliographystyle{apsrev.bst}
\pagenumbering{arabic}
\renewcommand{\figurename}{FIG.}
\def\tablename{TABLE}

\title{Comparative Study of Strain-Engineered Thermoelectric Performance of\\ 
2D-Xene Nanoribbons}

\author{Kalpana Panneerselvam, Swastik Sahoo and Bhaskaran Muralidharan}
\thanks{corresponding author: bm@ee.iitb.ac.in}
	
\affiliation{Department of Electrical Engineering, Indian Institute of Technology Bombay, Powai, Mumbai-400076, India}
\begin{abstract}
The quest for efficient and scalable thermoelectric materials has catalyzed intense interest in quasi-1D nanoribbons, where reduced dimensionality and structural tunability can decouple key transport parameters to enhance energy conversion. In this work, we present a unified comparative study of the thermopower in armchair nanoribbons derived from five archetypal 2D materials: graphene, silicene, germanene, stanene and phosphorene. Using a tight-binding model parametrized by first-principles inputs and solved within the  Landauer–B\"uttiker  formalism, we compute strain and width-dependent thermopower across nanoribbons classified by width families (3p, 3p+1, 3p+2) over a wide range of uniaxial tensile strain. Our results reveal that thermoelectric behavior is governed by a complex interplay of bandgap evolution, chemical potential asymmetry, and quantum confinement. While graphene and silicene exhibit pronounced family and width sensitive thermopower enhancement under moderate strain, heavier Xenes such as germanene and stanene show diminished responses. In particular, phosphorene nanoribbons emerge as exceptional, exhibiting remarkably high thermopower ($\approx \pm 62\,k_{B}/e$), a consequence of their large, persistent bandgap and anisotropic electronic structure. Across all systems, the 3p+2 family transitions from near-metallic to semiconducting under strain, enabling dramatic activation of thermopower in previously inactive configurations. This systematic cross-material analysis delineates the design principles for the optimization of TE in 1D nanoribbons, highlighting the strategic use of width control and strain engineering. Our findings identify phosphorene as an intrinsically superior thermoelectric material and position-strained Xene nanoribbons as promising candidates for tunable, low-dimensional thermoelectric devices.
\end{abstract}
\maketitle
\section{Introduction}
\indent The global demand for sustainable and efficient energy conversion technologies has intensified research into thermoelectric (TE) materials, which can directly convert heat into electricity via the Seebeck effect. TE devices provide a solid-state platform for energy harvesting, particularly for waste heat recovery in microelectronics, wearables, and autonomous systems \cite{snyder2008complex,he2017advances, zebarjadi2012perspectives}. The performance of a TE material is characterized by the dimensionless figure of merit $ZT = S^2 \sigma \:T\:/ \kappa$, where S is the thermopower, $\sigma$ is the electrical conductivity, T is the absolute temperature and $\kappa$ is the total thermal conductivity, comprised of electronic and lattice contributions \cite{goldsmid2010introduction}. Among these parameters, S plays a pivotal role in optimizing the power factor $PF = S^2 \sigma$, particularly in low-dimensional systems. However, achieving high ZT is challenging due to the interdependent nature of these parameters, enhancing $\sigma$ often leads to reduced S or increased $\kappa$ \cite{goldsmid2010introduction,bell2008cooling, shakouri2011recent}.\\
\indent A major breakthrough came with the prediction that low-dimensional materials can overcome these trade-offs via quantum confinement and boundary phonon scattering. Hicks and Dresselhaus showed that in 1D and 2D systems, the power factor can be enhanced while suppressing $\kappa$, boosting ZT \cite{hicks1993effect, hicks1993thermoelectric}. Since then, 2D materials and their quasi-1D nanoribbon derivatives have emerged as promising platforms for TE optimization due to their tunable band structures, high surface-to-volume ratios, and strong response to external perturbations such as strain($\varepsilon$) \cite{li2020recent, lv2014enhanced, chegel2023tunable, xuan20231d, zuev2009thermoelectric, nguyen2015enhanced}. Graphene nanoribbons (GNRs) provided the first experimental testbed for this vision. The opening of a width‑dependent band gap (0.1–2.2eV for $w < 5 nm$) \cite{han2007energy, son2006energy} together with tunable edge chemistry and external stimuli such as $\varepsilon$ and electric fields yields S in the range of 0.9–3.7 mV/K \cite{ouyang2009theoretical, zheng2012enhanced, liang2012enhanced, yeo2013first, yeo2012strain, liang2013electronic, vu2018enhancement}, an order of magnitude higher than in gapless graphene. However, the intrinsically high lattice thermal conductivity of graphene($\approx 2000\,W/mK$) remains a bottleneck to achieve a high ZT \cite{balandin2011thermal, hu2009thermal, jauregui2010thermal}. \\
\indent Elemental \enquote{Xenes}, buckled analogues of graphene such as silicene, germanene, and stanene, circumvent this limitation. Their mixed $sp^{2}-sp^{3}$ bonding breaks sublattice symmetry and endows pristine sheets with tunable gaps, sizeable spin–orbit coupling (SOC) and drastically lower $\kappa$ \cite{molle2017buckled, balendhran2015elemental, ezawa2014topological, ni2012tunable, drummond2012electrically, yan2015tuning, sahoo2022silicene, sahoo2024density}. When patterned into nanoribbons (SNRs, GeNRs, and StNRs), Xenes exhibit enhanced TE performance through phonon engineering \cite{sadeghi2015enhanced, cao2023significant, khan2017characterization}, edge passivation \cite{pan2012thermoelectric, monshi2017edge, navid2018thermal}, energy spectrum engineering through strain or gating \cite{zaminpayma2016band, gonzalez2023tuning, yang2014thermoelectric, xu2013large}, magnetic or topological effects \cite{zberecki2013thermoelectric,krompiewski2017edge, ildarabadi2021edge}, and large-scale atomistic modeling \cite{balatero2015molecular}. Phosphorene nanoribbons (PNRs) push the envelope even further. The puckered black phosphorus lattice imparts pronounced inplane anisotropy, direction-dependent band dispersion \cite{xia2014rediscovering, qiao2014high, taghizadeh2015scaling} and ultralow lattice thermal conductivities ($<1\,W/mK$), with S exceeding 1.3 mV/K, and reaching 2.6 mV/K after edge hydrogenation, supporting predicted ZT as high as 6.4 at room temperature \cite{zhang2014phosphorene}. Beyond elemental monolayers, transition-metal dichalcogenides and MXenes enrich the palette of materials, but challenges remain due to high $\kappa$ or metallic conduction \cite{nair2025ultra, kumar2024comparative, chen2023ambipolar, salah2024suppressing, huang2024violation}.\\  
\indent Despite these advances, most TE studies focus on single materials, lacking consistent cross-material comparisons. The reported values of S, $\sigma$, and ZT vary widely due to modeling assumptions, edge effects, or temperature ranges. Moreover, while $\varepsilon$ is a known tuning parameter, its impact on S in different families of nanoribbons is not well mapped. To address this, we present a unified comparative study of strain-tunable TE behavior in nanoribbons derived from five prototypical 2D Xenes: graphene, silicene, germanene, stanene, and phosphorene. Using a tight-binding model (TB) informed by hopping parameters derived from density functional theory (DFT) and evaluated within the Landauer–B\"uttiker (LB) formalism, we compute S as a function of the width of the ribbon and uniaxial tensile strain. By isolating S, a key electronic descriptor, we decouple the intrinsic response of TE from artifacts related to $\sigma$ and $\kappa$, providing a clearer view of the modulation induced by $\varepsilon$ in material systems.\\
\indent The structure of the paper is as follows. Section \ref{sec_method} outlines the mathematical framework employed for the simulations. In Section \ref{sec_results}, we present and analyze the results on the evolution of the strain-induced band structure modifications and the thermopower. Finally, Section \ref{sec_summary} summarizes the key findings and conclusions of the study.\\
\section{Theoretical Model}
\label{sec_method}
\begin{figure*}[!t]
    \centering
    \includegraphics[width=\textwidth,height=0.54\textheight]{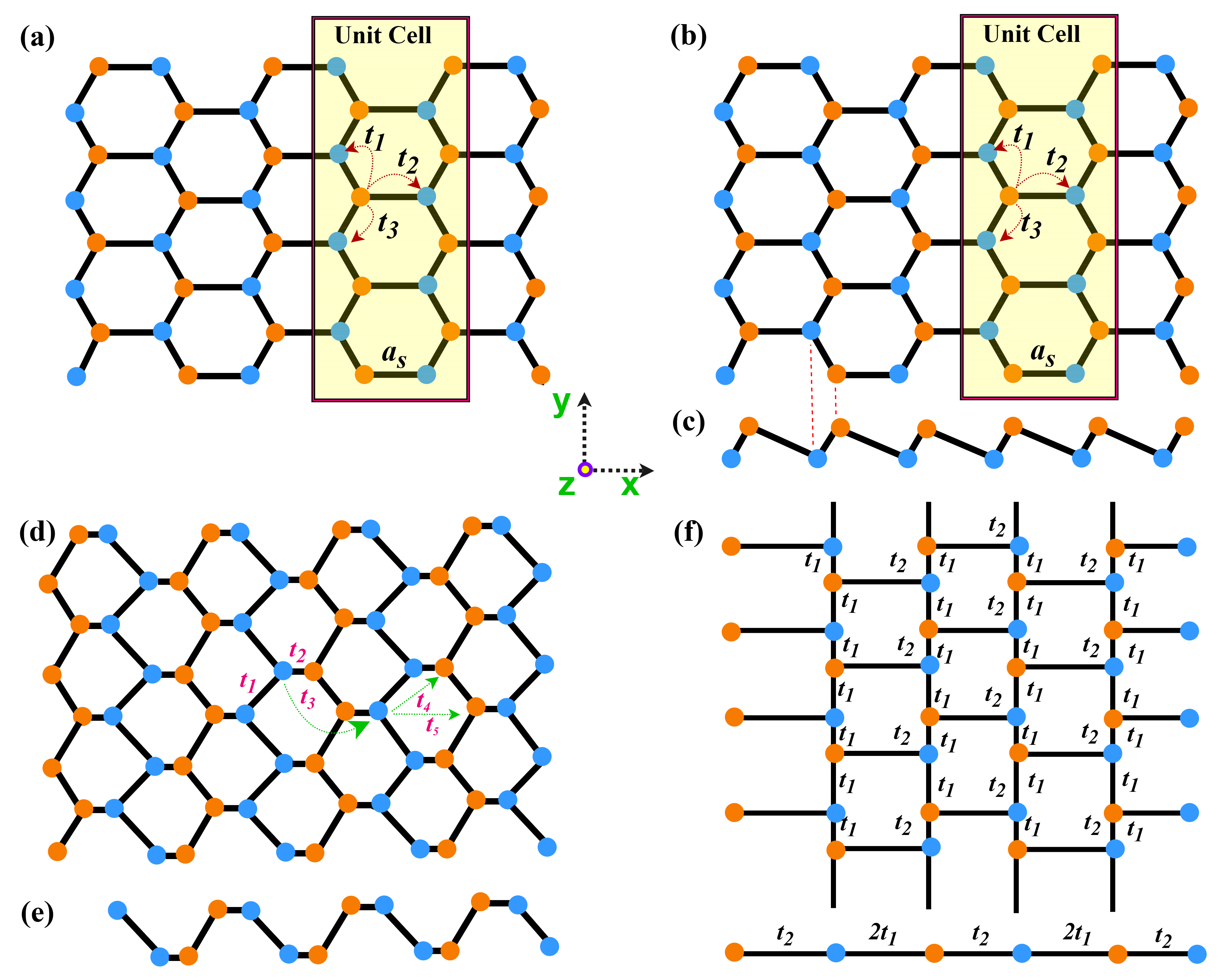}
    \caption{Top and side views of the atomic structure of N = 8 armchair-edge nanoribbons: (a) Graphene nanoribbon (GNR); (b) Top View, (c) Side view of Buckled Xene nanoribbons—Silicene, Germanene, and Stanene. (d) Top view, (e) Side view of Armchair phosphorene nanoribbon (PNR) with puckered lattice geometry, (f) Topologically equivalent representation of PNR modeled using a two-parameter tight-binding (TB) framework \cite{taghizadeh2015scaling}.}
    \label{lattice structure}
\end{figure*}
\indent The schematics of the lattice structures of the armchair nanoribbons derived from a family of $2D$ materials that include graphene, silicene, germanene, stanene and phosphorene are shown in Figs.~\ref{lattice structure}(a-f). Each of these nanoribbons exhibits distinct structural features that critically influence their electronic and TE properties. GNRs are planar, formed by cutting monolayer graphene along the armchair crystallographic direction, resulting in flat $sp^2$ hybridized carbon atoms arranged in a honeycomb lattice (Fig.~\ref{lattice structure}a). In contrast, Xene-based nanoribbons feature a buckled honeycomb lattice due to partial $sp^3$ hybridization (Fig.~\ref{lattice structure}b), which breaks the inversion symmetry and opens the bandgaps even in the absence of external fields. The degree of buckling increases with atomic number, which also enhances SOC in heavier Xenes such as stanene. PNRs display a puckered orthorhombic lattice (Fig.~\ref{lattice structure}d), leading to strong in-plane anisotropy and direction-dependent orbital overlap.\\
\begin{figure}[!b]
 \centering
    \includegraphics[width=\linewidth]{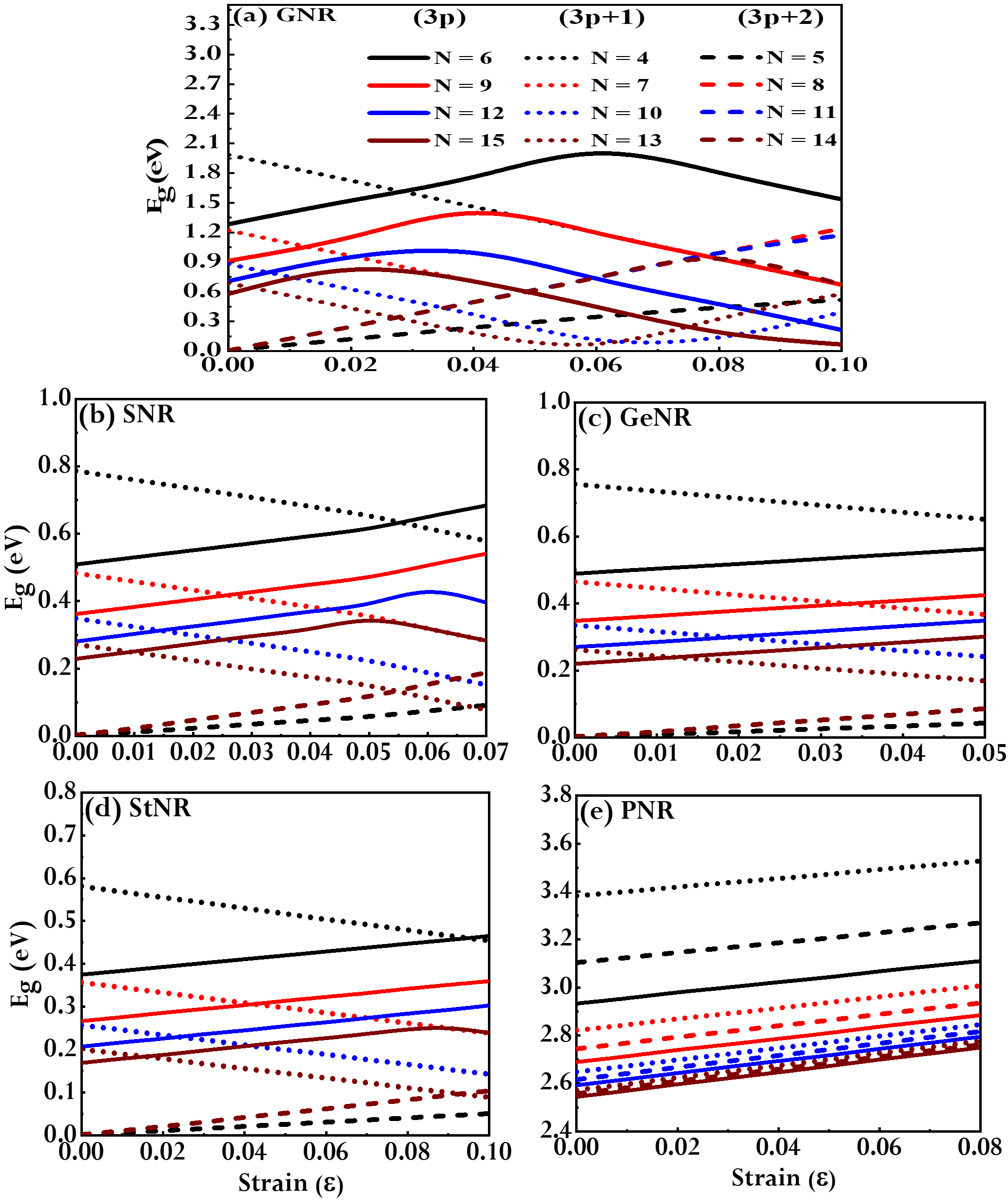}
    \caption{Evolution of electronic bandgaps ($E_{g}$) in N = 4–15 armchair-edge nanoribbons under $\varepsilon$ for (a) graphene (GNR), (b) silicene (SNR), (c) germanene (GeNR), (d) stanene (StNR), and (e) phosphorene (PNR). Ribbons are classified into width families: 3p (solid lines), 3p+1 (dotted lines), and 3p+2 (dashed lines). Each curve traces the strain dependence of the bandgap for a specific ribbon width, highlighting distinct family-dependent trends across different materials.}
    \label{fig:2}
  \end{figure}
\indent We focus exclusively on armchair-edge terminated nanoribbons, where the electronic transport is considered along the periodic (longitudinal) direction of the nanoribbon. Following standard convention, armchair graphene nanoribbons (AGNRs) are characterized by the number of dimer lines across their width and are denoted as N-AGNRs, where N specifies the ribbon width. For example, in GNRs, this width-dependent classification determines whether the ribbon is metallic or semiconducting, with periodicity in the bandgap behavior observed for N = 3p, 3p + 1 and 3p + 2, where p is an integer. A similar width-dependent classification applies to other Xene nanoribbons and PNRs. However, trends in bandgap scaling vary due to their buckled or puckered geometries and the influence of stronger SOC in heavier elements. These structural variations are crucial, as they control the confinement-induced bandgap, the density of states (DOS) near the Fermi level, and ultimately the TE performance.\\
\indent This study investigates the TE response of strain-engineered nanoribbons derived from 2D Xene nanoribbons using a unified TB and quantum transport framework. Our approach focuses exclusively on the S, computed from the electronic transmission spectrum under varying $\varepsilon$ and N. Each nanoribbon is modeled as a quasi-1D system with translational periodicity along the transport direction and a finite width defined by the number of atomic rows N in the transverse direction. Under $\varepsilon$, the length of the repeating unit cell becomes $ \ell =(1+\nu \varepsilon)\ell_{0}$ \cite{yeo2013first, sahoo2024density}, where $\ell_{0}$ is the length before deformation and $\nu$ is the Poisson ratio. The electronic structure is captured using a nearest-neighbor TB Hamiltonian of the form:
\begin{equation}
    H = \sum_{i} E_{i} c_i^{\dagger} c_i + \sum_{i,j} t_{ij} (\varepsilon) c_i^{\dagger} c_j + H.C.
    \label{eq:placeholder}
\end{equation}
\noindent where $c_{i(j)}^{\dagger}$ and $c_{i(j)}$ are the creation and annihilation operators at site i(j), $E_{i}$ is the on-site energy, and $t_{ij} (\varepsilon)$ is the strain-dependent hopping integral between neighboring sites i and j. $\varepsilon$ modifies the hopping amplitudes through the empirical scaling relation \cite{sahoo2022silicene}:
\begin{equation}
    t_{ij}(\varepsilon) = t_{ij}^0 \exp\left[-\beta_s\left(\frac{r}{r_0} - 1\right)\right]
\end{equation}
\noindent Here, $t_{ij}^0$  and $r_0$  are the unstrained hopping parameters and bond lengths, respectively, and $\beta_s$ is the decay constant. The strain-modified hopping parameters $t_{ij}(\varepsilon)$ are extracted from the DFT calculations reported by Sahoo et al. \cite{sahoo2024density}. For PNRs, the five-parameter model is reduced to two dominant terms ($t_{1}$ and $t_{2}$) using a topological equivalence approach \cite{taghizadeh2015scaling}, preserving essential characteristics near the Fermi level for transport analysis. The applied $\varepsilon$ is tailored for each material system based on its mechanical stability and prior studies \cite{sahoo2024density}: GNRs (0-10\%), SNRs (0-7\%), GeNRs (0-5\%), StNRs (0-10\%) and PNRs (0-8\%). To efficiently compute the electronic band structure and DOS, the Hamiltonian is recast in the matrix form suitable for periodic systems \cite{chuan2019electronic, datta2005quantum}:
\begin{equation}
    H(k) = \alpha_s + \beta_s \, e^{i k a_s} + \beta_s^* e^{-i k a_s}
    \label{eq:1}
\end{equation}
\noindent Here, $\alpha_s$ is the intra-cell (onsite) Hamiltonian matrix, $\beta_s$ is the inter-cell hopping matrix, k is the wave vector along the transport direction and $a_s$ is the lattice constant of the strained unit cell with $a_s = a_0 (1+\nu \varepsilon)$. Diagonalizing H(k) yields the energy eigenvalues $E_{n}(k)$, which form the band structure. Once strain-dependent TB parameters are established, electronic transport is analyzed within the LB formalism. Central to this approach is the calculation of the transmission function $\mathcal{T}(E)$, which encapsulates the energy-resolved probability that charge carriers will cross the nanoribbon. In the quasi-ballistic regime, where the scattering is minimal and phase coherence is preserved, the energy-dependent total transmission function $\mathcal{T}_{s}(E)$ of the strained system is expressed as the product of the transmission probability $\mathcal{T}(E)$ and the number of conducting modes (or mode density) M(E), which is formally defined as \cite{datta2005quantum},
\begin{equation}
    M(E) = \sum_n \sum_k \delta(E - E_{n,k}) \frac{1}{\ell_s} \frac{\partial E_{n,k}}{\partial k}
    \label{eq:placeholder_label}
\end{equation}
with $\delta(E - E_{n,k})$ is approximated by a Lorentzian broadening function \cite{datta2005quantum, chuan2019electronic}:
\begin{equation}
    \delta(E - E_{n,k}) = \frac{\eta / 2\pi}{(E - E_{n,k})^2 + (\eta/2)^2}
\end{equation}
where, $\eta$ is a very small positive value to prevent the inverse matrix from diverging. A large number of k points (typically 1000) is used for convergence. The band structures provide direct insight into the modulation of the band gap under $\varepsilon$ and the emergence of van Hove singularities near the band edges, which strongly influence the S. Within the linear response regime, the transmission probability can be written as \cite{sahoo2024density}:
\begin{equation}
  \mathcal{T}(E) = \frac{\lambda(E)}{\lambda(E) + \ell_s}
  \label{eq:example}
\end{equation}
where $\lambda (E)$ is the mean free path as a function of carrier energy. This expression effectively captures the impact of finite-size effects and coherence length on determining the energy-filtering efficiency of the channel. Using the calculated $\mathcal{T}_{s}(E)$, the S is evaluated from the LB formalism in the linear-response regime \cite{yeo2013first, ouyang2009theoretical, kuo2024thermoelectric}:
\begin{equation}
    S(\mu, T) = -1 / eT \times \left[ L_1(\mu, T) / L_0(\mu, T) \right]
    \label{eq:example}
\end{equation}
with the transport integrals $\mathcal{L}_n$ defined by:
\begin{equation}
    \mathcal{L}_n = \frac{2}{h} \int dE \, \mathcal{T}_{s}(E)(E - \mu)^n \frac{\partial f(E, \mu, T)}{\partial \mu}.
\end{equation}
\noindent where $\mu$ is the chemical potential (typically set at the Fermi level), $f(E, \mu, T)$ is the Fermi–Dirac distribution function, T is the temperature (in the present study all calculations are computed at room temperature), and e is the elementary charge. The integrals $L_0$ and $L_1$ capture, respectively, the contribution of electrical conductance and the asymmetry in transmission around the Fermi level.
\begin{figure}[!htbp]
  \begin{minipage}[!htbp]{0.48\textwidth}
    \centering
    \includegraphics[width=\textwidth]{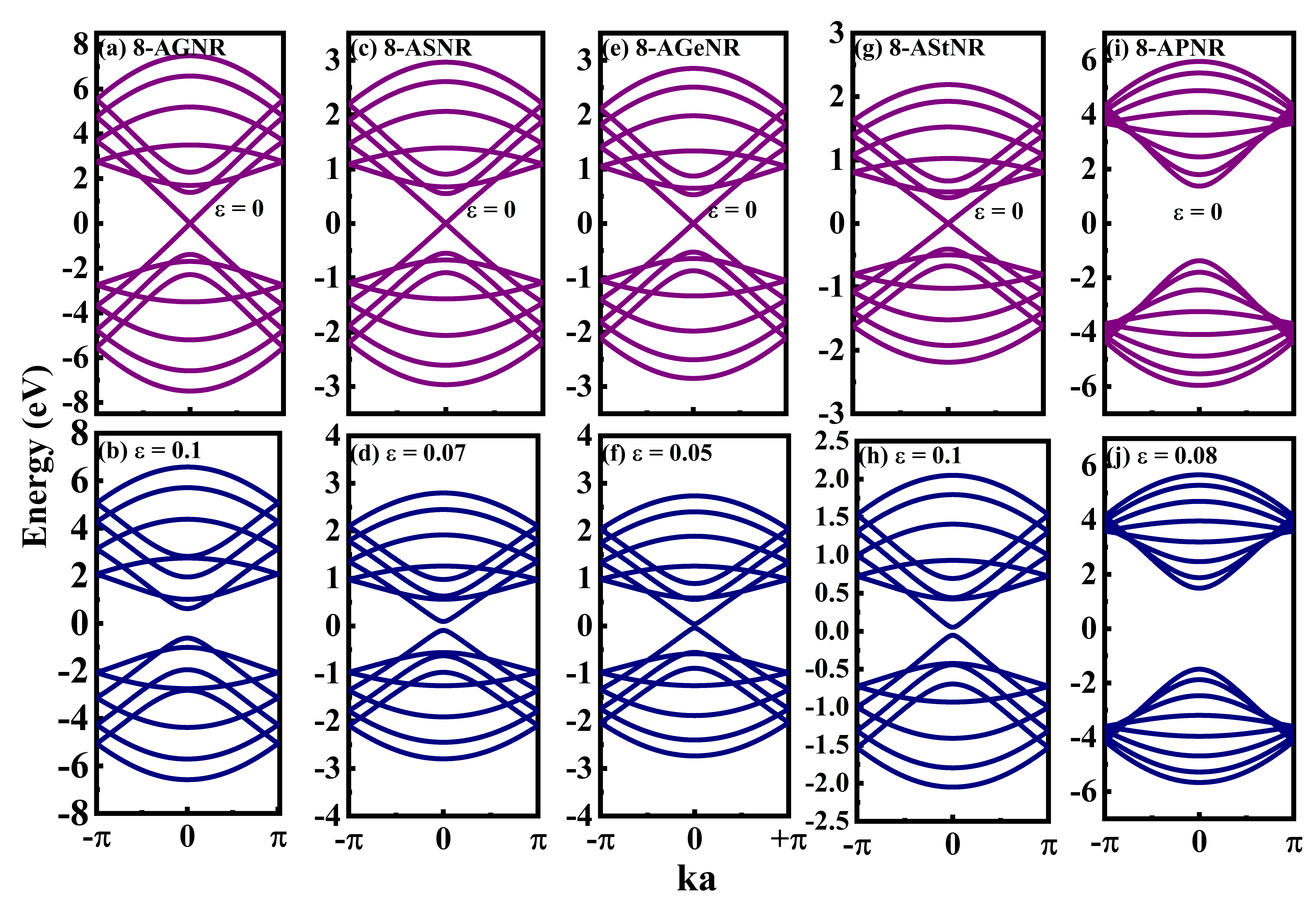}
    \caption{Strain-induced modulation of the electronic band structure in N = 8 armchair nanoribbons (3p+2 family) for different 2D materials. Top panels (a, c, e, g, i) show the band structures in the unstrained state ($\varepsilon$ = 0), while bottom panels (b, d, f, h, j) display the corresponding band structures under maximum applied $\varepsilon$: (a, b) AGNR, (c, d) ASNR, (e, f) AGeNR, (g, h) AStNR, and (i, j) APNR.}
    \label{fig:3}
  \end{minipage}\hfill\hspace{1cm}%
  \begin{minipage}[!htbp]{0.48\textwidth}
    \centering
    \includegraphics[width=\textwidth]{Fig._4.png}
    \caption{Strain-dependent number of conducting modes M(E) as a function of energy E for N = 8 armchair nanoribbons: (a) AGNR, (b) ASNR, (c) AGeNR, (d) AStNR, and (e) APNR. Black curves represent the unstrained condition ($\varepsilon = 0$), while orange curves denote the strained state at the maximum applied $\varepsilon$ for each material.}
    \label{fig:4}
  \end{minipage}
\end{figure}
\begin{figure*}[!htbb]
    \includegraphics[width=\linewidth, height=10cm]{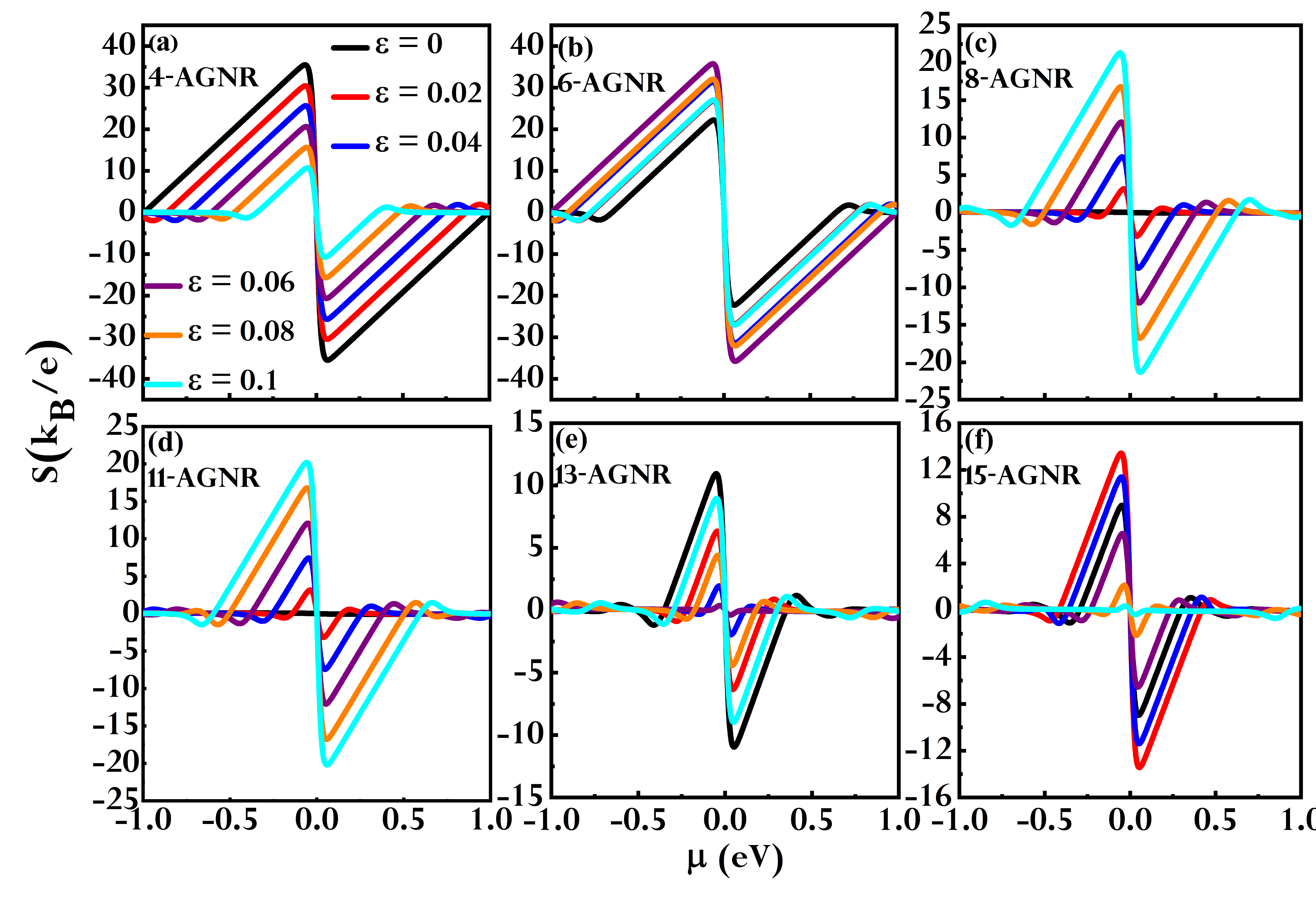}
    \caption{Strain-dependent thermopower S as a function of $\mu$ for AGNRs of different widths: (a) N = 4, (b) N = 6, (c) N = 8, (d) N = 11, (e) N = 13, and (f) N = 15. Each panel shows the evolution of $S(\mu)$ under varying uniaxial tensile strain values $\varepsilon = 0$ to $\varepsilon = 0.1$.}
	\label{Seebeck_GNR}
\end{figure*}\\
\indent The full modeling framework is developed in MATLAB using custom-built TB and transport solvers. For each material and the applied $\varepsilon$, the simulation is performed in a systematic way. First, the geometry and unit cell of the armchair nanoribbon are generated. Based on atomic connectivity and strain-modified hopping parameters, the matrices on site ($\alpha_s$) and inter-cell ($\beta_s$) are constructed. H(k), which depends on the wave vector, is then diagonalized to obtain the energy dispersion $E_n(k)$, from which the DOS is calculated. $\mathcal{T}_{s}(E)$ is evaluated directly from the band structure, and finally the S is calculated by integrating the transmission function within the LB formalism under the linear response regime. This modular and uniform approach ensures that all material systems are treated consistently, allowing for a direct and meaningful comparison of their strain-dependent TE behavior.
\section{Results and Discussion}
\label{sec_results}
\begin{figure*}[!htbp]
    \includegraphics[width=\linewidth, height=10cm]{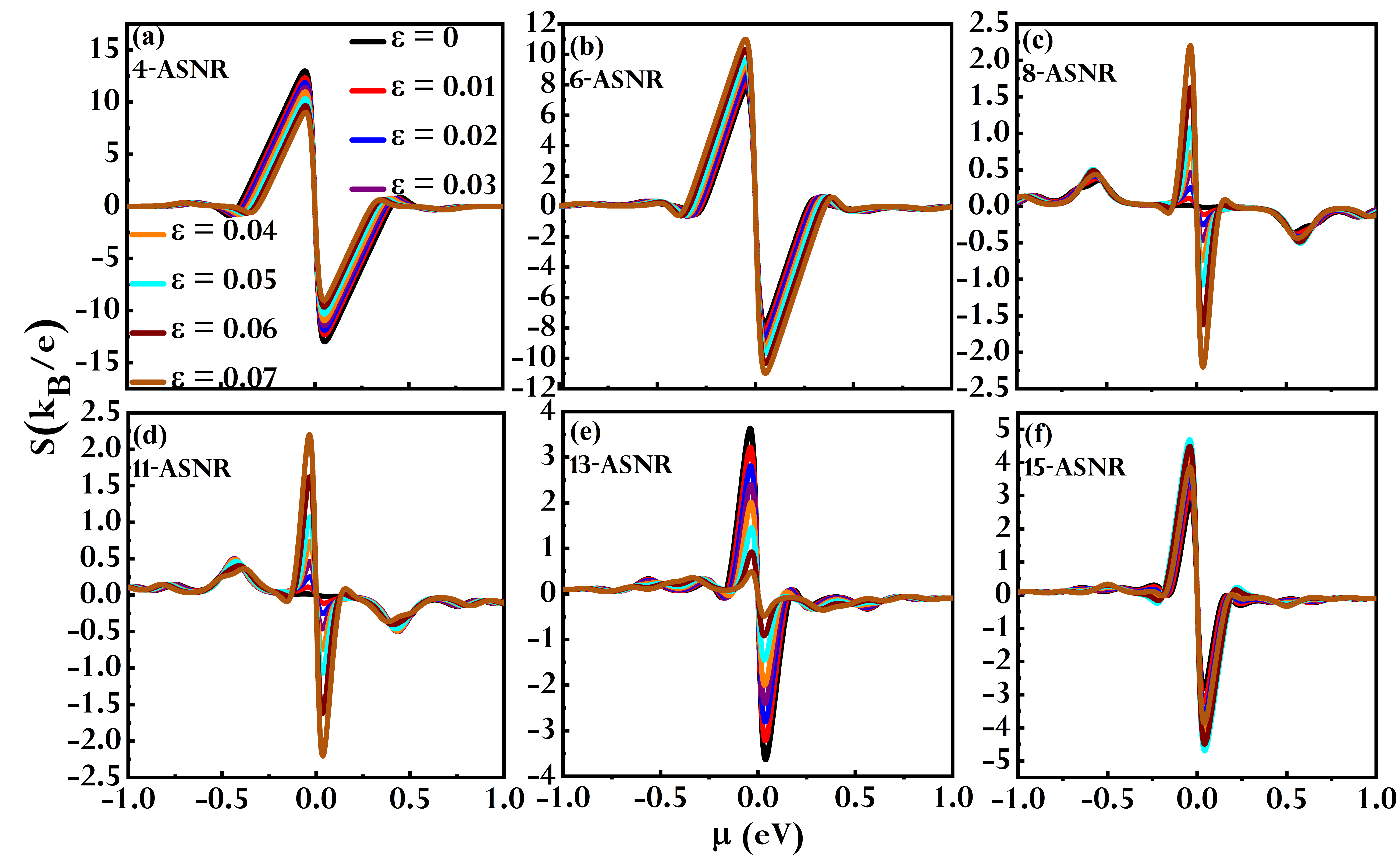}
    \caption{Strain-dependent thermopower S as a function of $\mu$ for ASNRs of different widths: (a) N = 4, (b) N = 6, (c) N = 8, (d) N = 11, (e) N = 13, and (f) N = 15. Each panel shows the evolution of $S(\mu)$ under varying uniaxial tensile strain values $\varepsilon = 0$ to $\varepsilon = 0.07$.}
	\label{Seebeck_SNR}
\end{figure*}
\begin{figure*}[!htbp]
    \includegraphics[width=\linewidth, height=10cm]{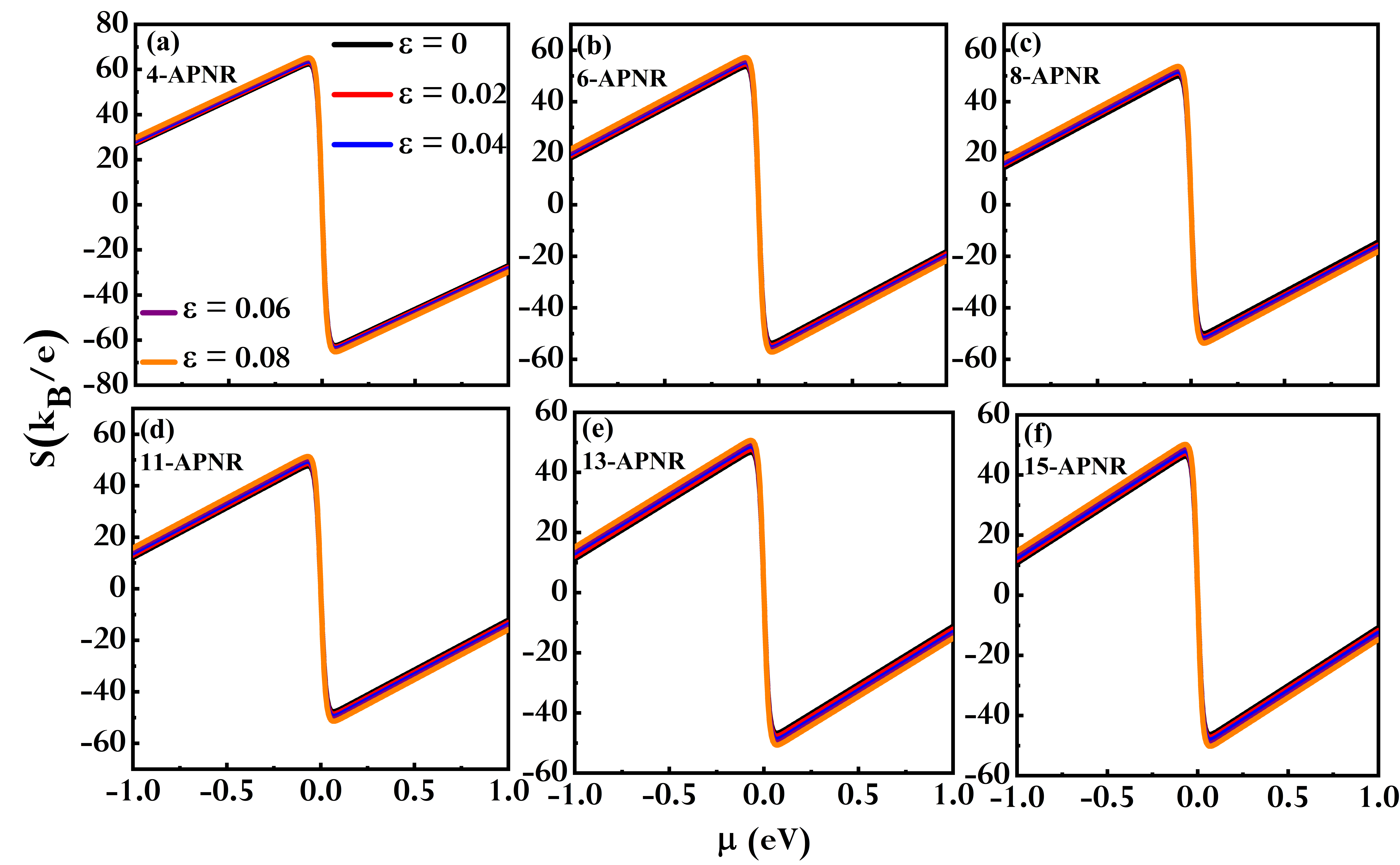}
    \caption{Strain-dependent thermopower S as a function of $\mu$ for APNRs of different widths: (a) N = 4, (b) N = 6, (c) N = 8, (d) N = 11, (e) N = 13, and (f) N = 15. Each panel shows the evolution of $S(\mu)$ under varying uniaxial tensile strain values $\varepsilon = 0$ to $\varepsilon = 0.08$.}
	\label{Seebeck_PNR}
\end{figure*}

\indent The evolution of electronic band gaps in nanoribbons derived from $2D$ materials have been systematically examined as a function of the width of the ribbon (N = 4 to 15) and $\varepsilon$, as shown in Figs.~\ref{fig:2}(a-e). Classifying these nanoribbons into the 3p, 3p+1, and 3p+2 width families reveals the significant role of quantum confinement in shaping their electronic properties, consistent with established theoretical models [Son et al., PRL 2006].\\ \\
\indent Graphene nanoribbons serve as a canonical example, displaying a rich width- and strain-dependent band structure (Fig.~\ref{fig:2}a). In the 3p family, the bandgap exhibits a non-monotonic trend as it increases with $\varepsilon$ up to a critical point ($\varepsilon \approx 0.06–0.08)$, beyond which it decreases. Both the gap magnitude and the optimal $\varepsilon$ at which the peak occurs decrease with increasing ribbon width. This behavior results from the competing effects of enhanced quantum confinement and strain-induced $\pi$ orbital realignment. In contrast, the 3p+1 family shows a strong width dependence. Narrower ribbons undergo a monotonic gap reduction under $\varepsilon$, while wider ribbons, first decreasing, then increasing beyond moderate $\varepsilon$. The 3p+2 family is nearly metallic at zero $\varepsilon$, but undergoes a steady and monotonic opening of the bandgap under $\varepsilon$, saturating at high strain levels. This strain activation makes the 3p+2 family particularly relevant for improving TE.\\ 
\indent Buckled Xenes exhibit qualitatively similar family-dependent trends (Fig.~\ref{fig:2}(b-d)). For these materials, the 3p and 3p + 2 families, respectively, show monotonic gap increases and strain-activated gap openings, while the 3p+1 family sees a steady decrease in bandgap with $\varepsilon$. However, the absolute gap values and the sensitivity to $\varepsilon$ vary by material. Narrow ribbons across all families exhibit larger bandgaps and greater tunability as a result of stronger confinement effects. Among Xenes, SNRs demonstrate the highest strain tunability, particularly in the 3p + 2 family, reaching $\sim0.25eV$ at $\varepsilon = 0.10$. GeNRs follow similar trends with slightly lower magnitudes. StNRs show the weakest response; even under maximum strain, bandgaps generally remain below 0.15eV, attributed to the heavier atomic mass and stronger SOC. While non-monotonic features are prominent in GNRs, especially in wider members of the 3p and 3p+1 families, they are largely absent in Xenes. For silicene, germanene, and stanene, the gap evolution with $\varepsilon$ is mostly linear or gently curved. Only at high strain and large width do minor plateaus or inflections appear. This linearity is likely due to stronger $\sigma–\pi$ orbital mixing, structural buckling, and reduced edge localization compared to flat GNRs.\\
\indent Armchair PNRs distinguish themselves by exhibiting a consistent and predictable electronic response across all width families. They retain large semiconducting bandgaps, for instance, ranging from 3.37eV at N = 4 to 2.61eV at N = 13, with the gap decreasing monotonically as the width of the ribbon increases, in agreement with previous theoretical predictions \cite{taghizadeh2015scaling, zhang2014phosphorene, naemi2019modulation}. Under applied $\varepsilon$, all PNRs show a nearly linear increase in bandgap, regardless of their width family, as shown in Fig.~\ref{fig:2}e. In particular, no metallic transitions or non-monotonic behavior are observed, highlighting the exceptional structural and electronic stability of the material, which is advantageous for device applications.\\ \\
\indent To further elucidate the effects of $\varepsilon$ on electronic properties, Fig.~\ref{fig:3} presents the band structures of N = 8 armchair nanoribbons for each material, both in the unstrained and maximally strained states. The 3p+2 family is emphasized, known for its metallic nature at zero $\varepsilon$ and strain-induced semiconductor transitions. In 8-AGNR (Figs.~\ref{fig:3}(a,b)), $\varepsilon = 0.1$ opens a distinct bandgap where the valence and conduction bands initially touch, confirming a metal-to-semiconductor transition. Similar trends are observed in ASNR, AGeNR, and AStNR (Figs.~\ref{fig:3}(c–h)), where small bandgaps increase with strain, and the dispersion of the subband becomes more pronounced near the Fermi level. APNRs (Figs.~\ref{fig:3}(i,j)) are already semiconducting at $\varepsilon=0$, the applied $\varepsilon$ further increases the bandgap modestly. The bands remain parabolic and symmetric, indicating robustness in deformation \cite{naemi2019modulation}. Collectively, these plots reinforce the critical role of $\varepsilon$ in tuning the electronic properties of 3p+2 nanoribbons, particularly for TE and switching applications.\\
\begin{figure*}[!htbp]
    \includegraphics[width=\linewidth, height=10cm]{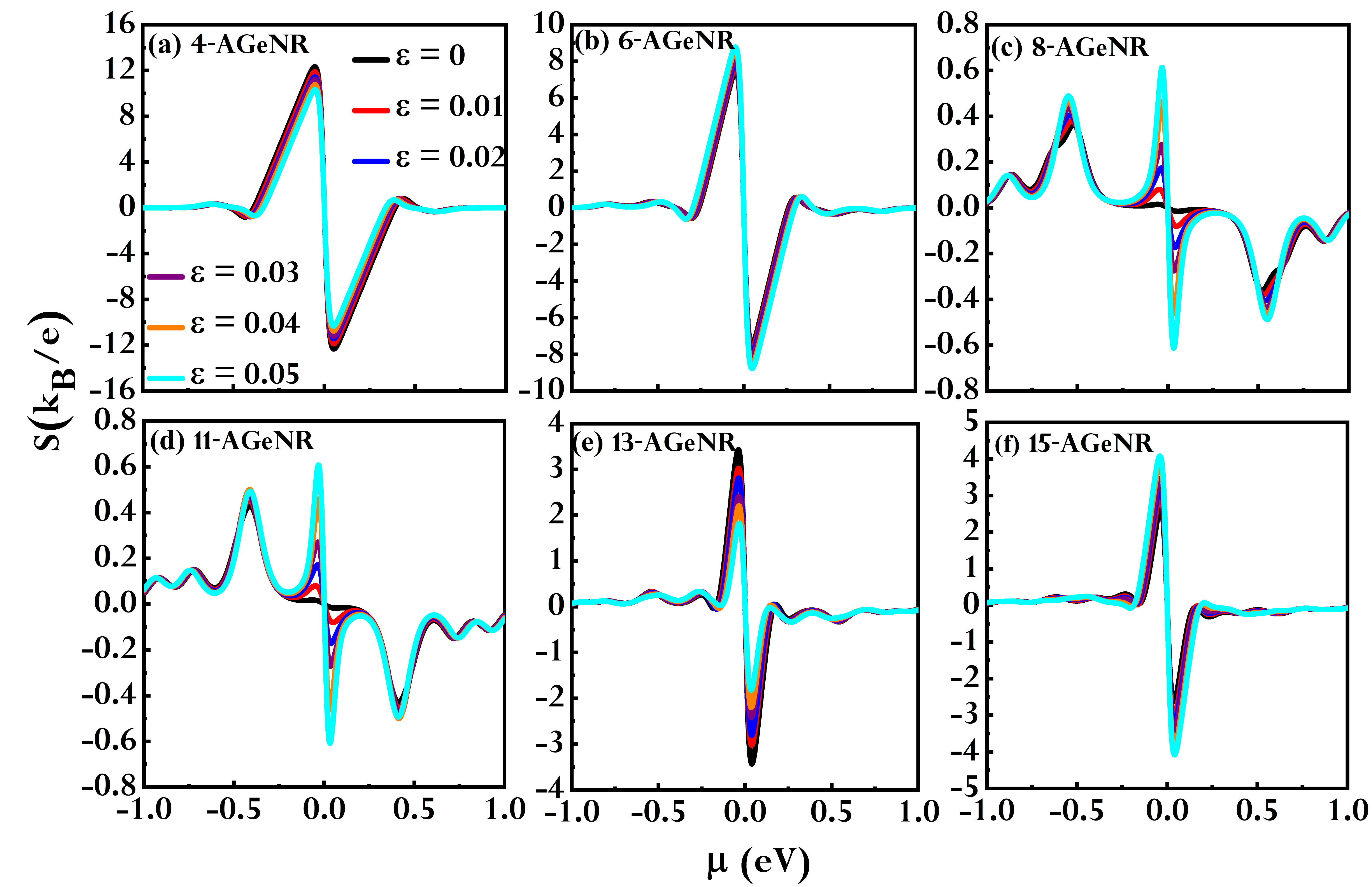}
    \caption{Strain-dependent thermopower S as a function of $\mu$ for AGeNRs of different widths: (a) N = 4, (b) N = 6, (c) N = 8, (d) N = 11, (e) N = 13, and (f) N = 15. Each panel shows the evolution of $S(\mu)$ under varying uniaxial tensile strain values $\varepsilon = 0$ to $\varepsilon = 0.05$.}
	\label{Seebeck_GeNR}
\end{figure*}
\indent To complement the band structure and thermopower analyses, we examine the energy resolved M(E) for the N = 8 armchair nanoribbons of the five 2D materials, as shown in Fig.~\ref{fig:4}. These profiles highlight how $\varepsilon$ modifies the availability of transport channels near the Fermi level and reveals the onset of conduction through quantum subbands. In the case of AGNRs (Fig.~\ref{fig:4}a), the unstrained system shows a near-metallic behavior with a small transport gap, and the M(E) profile displays a staircase-like progression typical of 1D quantization. Under a tensile deformation of $\varepsilon = 0.1$, a clear suppression of modes is observed near the Fermi level, with M(E) reducing to zero at E = 0. This indicates the opening of a bandgap and the transition to a semiconducting regime. The shift of the lowest plateau to higher energies reflects the strain-induced reconfiguration of the band edges. For ASNRs (Fig.~\ref{fig:4}b), strain induces a similar suppression of the mode and opening of the gap as in AGNRs, but with a smaller magnitude, reflecting a more modest reconfiguration of the transport channels. At $\varepsilon = 0.07$, the number of modes near the Fermi level further decreases, indicating the enhancement of the pre-existing bandgap due to $\varepsilon$. The smoother mode steps suggest increased subband separation and sharper density of states at band edges, promoting enhanced TE performance. In AGeNRs and AStNRs (Fig.~\ref{fig:4}c and ~\ref{fig:4}d), the unstrained ribbons already exhibit substantial bandgaps, with nearly vanishing modes at E = 0. Upon application of $\varepsilon = 0.05, and   0.1$, respectively, the mode profiles show a moderate widening of the gap region and a slight upward shift in the onset energy of the first step. However, the high-energy features remain largely unchanged, indicating that the strain primarily affects the low-energy conduction pathways. Notably, the smoother M(E) plateaus in these Xenes are attributed to the buckled lattice geometry and reduced confinement compared to that in graphene. In APNRs (Fig.~\ref{fig:4}e), M(E) is zero over a wide energy window even in the unstrained state, consistent with the large intrinsic bandgap of phosphorene. Under $\varepsilon = 0.08$, the width of the gap increases slightly, shifting the onset of the mode outward, but without introducing significant asymmetry. The robustness and symmetry of the M(E) profile highlight the structural and electronic resilience of phosphorene under strain, corroborating its geometry-insensitive TE behavior. Across the five systems, a key signature of strain is the downward shift or widening of the zero-mode region, particularly prominent in AGNRs and ASNRs. This reflects a suppression of low-energy conducting channels, thus increasing carrier energy selectivity, an essential criterion for enhancing the Seebeck coefficient. Additionally, the step-like features in M(E) become more distinct under strain in systems with initially small or no bandgaps, suggesting improved quantization and band separation due to strain-induced modifications in electronic structure.\\
\begin{figure*}[!htbp]
    \includegraphics[width=\linewidth, height=10cm]{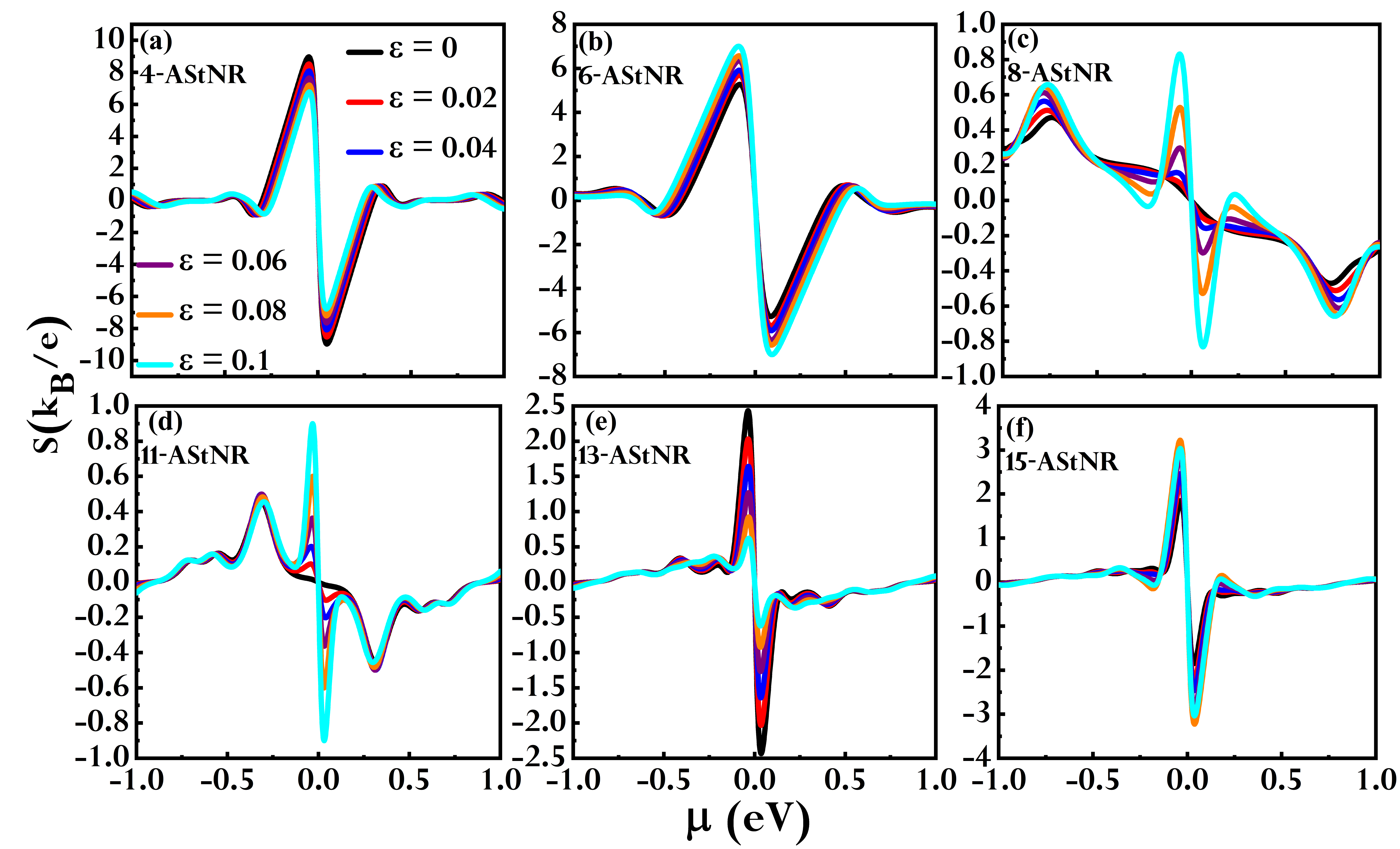}
    \caption{Strain-dependent thermopower S as a function of $\mu$ for AStNRs of different widths: (a) N = 4, (b) N = 6, (c) N = 8, (d) N = 11, (e) N = 13, and (f) N = 15. Each panel shows the evolution of $S(\mu)$ under varying uniaxial tensile strain values $\varepsilon = 0$ to $\varepsilon = 0.1$.}
	\label{Seebeck_StNR}
\end{figure*}
\indent To meaningfully interpret the TE behavior of the armchair nanoribbons, we examine S as a function of $\mu$ under varying widths and $\varepsilon$. This quantity reflects the asymmetry in carrier transport and is highly sensitive to band structure modifications induced by quantum confinement and mechanical deformation. We use a specific constant for the unit of thermopower S: $k_{B}/e = 86.25\,\mu V/K$. Given the extensive data set covering the widths of the ribbons N = 4 to N = 15 in three width families and multiple strain levels, we present a carefully curated subset of results. For each material system, two widths of the ribbons were chosen from each family classification (3p, 3p + 1 and 3p + 2): one in the narrow width regime (e.g. N = 4, 6, or 8) and one in the wider regime (e.g. N = 11, 13, or 15). This approach captures the key effects of quantum confinement and width scaling across different families, ensuring both comprehensive insight and clarity of presentation. Figures~\ref{Seebeck_GNR} to \ref{Seebeck_PNR} represent the computed thermopower profiles for AGNRs, ASNRs, AGeNRs, AStNRs and APNRs respectively. These graphs illustrate how the TE response evolves with $\mu$, N, and $\varepsilon$. However, the discussion refers to the full dataset where needed to highlight nuanced variations across all widths. In all cases, $S(\mu)$ is antisymmetric about $\mu = 0$, reflecting intrinsic electron–hole symmetry. However, peak values, sharpness, and the active range of $\mu$ vary significantly between materials and the family of ribbons. A universal feature observed in all materials is that the thermopower peaks do not emerge precisely at the band edges, but at $\mu$ that shift a few $k_{B}T$ from the midgap, where the carrier asymmetry in transport is maximized.\\
\indent In GNRs, quantum confinement is particularly strong for ultra-narrow ribbons (N = 4), where $S(\mu)$ exhibits high and broad peaks (up to $\pm 35\,k_{B}/e$ at $\mu \approx 0.065eV)$. The window of active $\mu$, defined as the range where $\mid S(\mu)\mid$ is significant, can span up to 0.25–0.35eV, reflecting a robust TE response over a wide doping range. This is a direct consequence of the large bandgap in narrow ribbons, which creates a strong electron-hole transport asymmetry. As the width increases, the bandgap shrinks, resulting in a sharp decrease in both the peak $S(\mu)$ and the width of the active window, which falls below 0.10eV for $N > 10$. This reduction is a direct manifestation of bandgap narrowing and the approach to bulk-like symmetry. The SNRs show similar qualitative family trends but with lower $S(\mu)$ peaks (up to $\pm 13\,k_{B}/e$ for N = 4), due to the buckled lattice and partial $sp^2–sp^3$ hybridization, which weakens the quantum confinement. With increasing width, both the amplitude and width of the thermopower peaks diminish further. GeNRs show even greater suppression: only the narrowest GeNRs (N = 4) display moderate $S(\mu)$ maxima ($\pm 12\,k_{B}/e$), and for all wider ribbons, the amplitude collapses rapidly, with the active window shrinking to below 0.05eV for $N \geq 8$. StNRs represent the extreme: the highest $S(\mu)$ is observed for N = 4 ($\pm 9\,k_{B}/e$), but wider ribbons are nearly thermoelectrically inactive due to small intrinsic bandgaps and strong SOC. For all widths, the active window is exceedingly narrow. A unified trend emerges for the 3p+2 family (N = 8, 11) in all Xenes: at zero strain, these ribbons are nearly metallic with vanishing or extremely small bandgaps, and thus exhibit uniformly suppressed $S(\mu)$, independent of the parent material. The absence of a significant bandgap eliminates the necessary carrier asymmetry for high thermopower, and $S(\mu)$ remains flat and close to zero for all relevant chemical potentials and ribbon widths. This “inactive” TE character is robustly confirmed by both the thermopower and bandgap data. PNRs stand out sharply from these trends. Here, $S(\mu)$ is exceptionally large up to $\pm 62\,k_{B}/e$ for N = 4 and remains substantial (still above $\pm 45\,k_{B}/e$) even for the widest ribbons studied (N = 15), far exceeding the values in Xenes. This is attributed to the anisotropic puckered lattice of phosphorene and the large, persistent bandgap ($E_{g} > 2.5eV$), which is robust in all widths and families. In particular, there is no family-dependent suppression: The 3p, 3p + 1, and 3p + 2 PNR all maintain strong, broad thermopower, establishing phosphorene as a unique geometry-insensitive, high-performance TE material.\\
\indent The evolution of $S(\mu)$ under uniaxial tensile deformation in AGNRs shows a sophisticated interaction between the width of the ribbon, the classification of the family, and the quantum-mechanical modification of the structure of the band. Detailed inspection of the strain-dependent $S(\mu)$ profiles reveals not only family and width specific trends in the magnitude of thermopower but also striking differences in the active $\mu$ window. For the 3p family, the effect of $\varepsilon$ on thermopower is non-monotonic and width-dependent. The narrow ribbons (N = 6, 9) exhibit a monotonic increase in $S(\mu)$ with $\varepsilon \approx 0.06$, after which further $\varepsilon$ leads to a decline. In wider ribbons (N = 12, 15), the optimal threshold for $\varepsilon$ decreases (to $\varepsilon \approx 0.04$ and $\varepsilon \approx 0.02$, respectively), marking an earlier onset of thermopower suppression with increasing $\varepsilon$. In addition to this behavior, the window of active $\mu$ for N = 6 expands from approximately 0.25eV at $\varepsilon=0$ to approximately 0.35eV at optimal $\varepsilon$ ($\varepsilon \approx 0.06$), indicating not only a higher thermopower but also a greater tolerance to fluctuations in $\mu$. For N = 12 and 15, while the maximum value of $S(\mu)$ is lower, the window still broadens at moderate $\varepsilon$ but contracts rapidly when the system is overstrained, mirroring the non-monotonic trend at peak $S(\mu)$. In the 3p+1 family, the $\varepsilon$ response divides into two distinct regimes. For narrower ribbons (N = 4, 7), the S shows a monotonic decrease as $\varepsilon$ is applied, with both the peak value and the width of the active $\mu$ window shrinking. For example, the window where $\mid S(\mu) \mid$ is substantial for N = 4 reduces from approximately 0.20eV at $\varepsilon=0$ to less than 0.10eV at $\varepsilon \approx 0.10$, highlighting a narrowing of operational flexibility. In wider ribbons (N = 10, 13), $S(\mu)$ decreases with strain up to $\varepsilon \approx 0.06$, after which both the peak value and the width of the high $S(\mu)$ region begin to recover, a reflection of strain-induced bandgap reopening. However, for all members of this family, the active window $\mu$ is generally narrower and more sensitive to applied $\varepsilon$ than in the other families, indicating that the 3p + 1 ribbons offer the least robust TE performance under mechanical deformation. In contrast, the 3p + 2 family displays a predominantly activating strain response. The ribbons in this family are nearly metallic at $\varepsilon=0$, with flat, negligible $S(\mu)$ profiles and a vanishing active $\mu$ window. As $\varepsilon$ increases, a bandgap opens and $S(\mu)$ increases rapidly and monotonically for N = 5, 8, and 11. The window of active $\mu$ expands from near zero at $\varepsilon = 0$ to as much as 0.28eV at $\varepsilon = 0.10$, a dramatic widening that enables practical device operation under a wide range of doping or gating conditions. For the widest 3p+2 ribbon (N = 14), this window and the peak value both increase up to $\varepsilon = 0.08$ and then contract for higher strains, again underscoring that overstraining can eventually suppress the favorable electronic asymmetry.\\
\indent The evolution of S under uniaxial $\varepsilon$ in silicene, germanene, and stanene nanoribbons demonstrates the remarkable tunability of quantum-confined 2D materials with respect to both ribbon width and mechanical deformation. All three Xenes exhibit pronounced family-dependent behavior, although the specific trends, rates, and achievable magnitudes of the strain response are material-dependent, highlighting the need for careful material selection and structural control in optimizing TE devices. SNRs (Fig.~\ref{Seebeck_SNR}) combine high-attainable S with complex width-dependent strain responses. In the 3p family (N = 6, 9, 12, 15), the narrow ribbons show a monotonic increase in $S(\mu)$ with $\varepsilon$, with peak values increasing from $7–9\,k_{B}/e$ at $\varepsilon=0$ to $10–12\,k_{B}/e$ at moderate strain ($\varepsilon \approx 0.06$). The window of active $\mu$, defined as the range where $\mid S(\mu) \mid$ exceeds $80\%$ of its peak, expands with $\varepsilon$, improving operational flexibility. For larger widths (N = 12, 15), this enhancement is limited to moderate $\varepsilon$, after which both the thermopower peak and the window decrease, reflecting a non-monotonic trend. In the 3p+1 family (N = 4, 7, 10, 13), both the peak $S(\mu)$ and the active window decrease monotonically with $\varepsilon$, reflecting a bandgap closing effect. The 3p+2 family (N = 5, 8, 11, 14) shows a monotonic increase in $S(\mu)$ with $\varepsilon$ for all widths: at $\varepsilon=0$, these ribbons are nearly metallic and inactive, but $\varepsilon$ opens a gap and activates significant thermopower. GeNRs (Fig.~\ref{Seebeck_GeNR}) present a fully monotonic relationship between $\varepsilon$ and thermopower across all families and widths, simplifying strain engineering. In the 3p family, $S(\mu)$ increases monotonically with $\varepsilon$ for all widths. In the 3p+1 family, $S(\mu)$ decreases monotonically, with the operating window contracting under strain. The 3p+2 family also shows a monotonic increase in $S(\mu)$ with strain, although both peak values and active windows are smaller than in silicene. StNRs (Fig.~\ref{Seebeck_StNR}) follow the same monotonic $\varepsilon$ trends across all families, but show the lowest S and the narrowest operational windows, due to weaker quantum confinement and pronounced suppression from heavy atomic mass and strong SOC. Silicene outperforms germanene and stanene in both the maximum S and the operating window, though each material shows distinct nuances under $\varepsilon$.\\
\indent Among all nanoribbon systems, PNRs are unique in showing a dominant width effect on thermopower, while $\varepsilon$ has a comparatively minor influence (Fig.~\ref{Seebeck_PNR}). As already explained, the maximum $\mid S(\mu) \mid$ values for narrow ribbons (N = 4 to 6) are exceptionally high ($\approx \pm 62\,k_{B}/e$), a direct reflection of strong quantum confinement and the intrinsically large bandgap of narrow APNR. As the width of the ribbon increases (N = 8, 10, 12, 15), the S decreases almost monotonically, dropping to around $46\,k_{B}/e$ at N = 15, in line with the systematic narrowing of the bandgap. In contrast, $\varepsilon$ causes only modest changes in thermopower for any given width: across all APNRs, the maximum $\mid S(\mu) \mid$ increases by at most $4-5\,k_{B}/e$ as $\varepsilon$ increases from zero to 0.08, a much smaller modulation than the intrinsic width dependence. The active $\mu$ window also shifts only slightly with $\varepsilon$, with no substantial broadening or narrowing, as observed in Xene nanoribbons. The relative insensitivity of APNRs to $\varepsilon$, despite their large thermopower, suggests that the band structure of phosphorene is already highly optimized for TE performance in its pristine form. Moderate uniaxial strain serves mainly as a fine-tuning mechanism rather than causing a fundamental change. The direct single-particle gap of phosphorene is less susceptible to closure or distortion under tensile deformation compared to that of the more flexible Xenes, likely because of its puckered structure and stronger in-plane anisotropy.
\section{Conclusions}
\label{sec_summary}
\indent Using a unified TB and Landauer formalism, we mapped how ribbon width, edge family, and uniaxial strain shape the thermopower response of armchair nanoribbons derived from graphene, silicene, germanene, stanene, and phosphorene. Among all the materials discussed in this paper, the strain provides the strongest leverage to GNRs. Among the buckled Xenes, silicene mirrors graphene behavior, delivering the largest strain‑driven gains within its cohort, while germanene yields steadier, yet smaller improvements, and stanene remains largely inert. The width and family-dependent variations in the active $\mu$ window across all materials are not just a technical detail, but have significant implications for the device. For the 3p and 3p+2 families, moderate tensile strain simultaneously increases the peak thermopower and broadens the $\mu$ window that maintains high S, imparting tolerance to the drift of Fermi levels from doping or environment; in contrast, 3p + 1 ribbons experience declining peaks and shrinking windows, limiting their suitability for flexible devices.\\ 
\indent In all three Xenes, the 3p+2 family demonstrates the most pronounced strain activation of thermopower; the 3p family offers strong, but width and strain-sensitive performance; and the 3p+1 family is consistently suppressed by strain. These trends emphasize the importance of careful family, width, and strain selection to optimize the TE performance of Xene-based nanoribbons. PNRs lie at the opposite extreme: quantum confinement alone pushes their thermopower to outstanding values that depend almost entirely on width, rendering strain comparatively ineffective and endowing these ribbons with intrinsic mechanical tolerance. These material‑specific responses lead to clear design rules.  Devices that require dynamic tunability should employ graphene or silicene ribbons from the 3p or 3p + 2 families, where the applied tensile strain maximizes both the height and breadth of the TE plateau. Where maximum, nearly strain‑insensitive thermopower is paramount, narrow phosphorene ribbons are the material of choice, with width control during synthesis replacing active strain engineering.  By matching the ribbon family, width, and strain to the target operating regime, the guidelines distilled here chart a direct route to high efficiency, atomically thin TE modules.\\
\bibliography{reference}
\end{document}